\documentclass[conference]{IEEEtran}
\IEEEoverridecommandlockouts

\usepackage{cite}
\usepackage{amsmath,amssymb,amsfonts}
\usepackage{algorithmic}
\usepackage{graphicx}
\usepackage{algorithm}
\usepackage{booktabs}
\usepackage{multirow}
\usepackage{textcomp}
\usepackage{xcolor}
\usepackage{eso-pic}
\def\BibTeX{{\rm B\kern-.05em{\sc i\kern-.025em b}\kern-.08em
    T\kern-.1667em\lower.7ex\hbox{E}\kern-.125emX}}
\begin{document}

\AddToShipoutPictureFG*{%
  \AtPageLowerLeft{%
    \put(\LenToUnit{0.625in},\LenToUnit{0.42in}){%
      \parbox[b]{\dimexpr\paperwidth-1.25in\relax}{%
        \footnotesize\raggedright
        This work has been submitted to the IEEE for possible publication.
        Copyright may be transferred without notice, after which this
        version may no longer be accessible.%
      }%
    }%
  }%
}

\title{Sparse Graph Neural Networks With Spectral Virtual Nodes for MIMO Detection\\
}

\author{
\IEEEauthorblockN{
Fatemeh Jalili and Ghaya Rekaya-Ben Othman
}
\IEEEauthorblockA{
LTCI, T\'el\'ecom Paris, Institut Polytechnique de Paris, Palaiseau, France\\
\{fatemeh.jalili, ghaya.rekaya\}@telecom-paris.fr
}
}

\maketitle

%%%%%%%%%%%%%%%%%%%%%%%%%%%%%%%%%%%%%%%%%%%%%%%%%%%%%%%%%%%%%%
\begin{abstract}
Graph neural network (GNN) detectors achieve high accuracy in large
multiple-input multiple-output (MIMO) systems, but exchange messages
between all pairs of transmitted symbols, so their complexity grows
quadratically with the number of streams. We first propose a sparse GNN
(SGNN) detector that restricts each symbol's messages to its most strongly
coupled neighbors, so the message count grows linearly with the system
size. However, pruning discards small couplings that jointly shape the
weak channel directions, which limit detection accuracy. Even on the dense
graph, learned messages capture these directions only partially. We
therefore propose a sparse GNN with spectral virtual nodes
(SVGNN), which filters the node states through a scaled regularized
inverse of the channel Gram matrix to emphasize the weak directions. With
two neighbors per symbol, SVGNN reduces the number of multiplications by a
factor of 18 relative to the approximate message passing GNN (AMP-GNN) in
a $64\times64$ MIMO system. Simulations with QPSK, 16-QAM, and 64-QAM over
i.i.d.\ and correlated channels show that SVGNN matches or outperforms
both AMP-GNN and expectation propagation, with gains of up to 3.2~dB, and
lowers the error floors of AMP-GNN in several configurations. The virtual
nodes improve AMP-GNN even without pruning, indicating that message
passing alone, whether sparse or dense, cannot fully capture the weak
directions. 
\end{abstract}

%%%%%%%%%%%%%%%%%%%%%%%%%%%%%%%%%%%%%%%%%%%%%%%%%%%%%%%%%%%%%%

\section{Introduction}
\label{sec:introduction}

Multiple-input multiple-output (MIMO) systems increase spectral efficiency
by transmitting several streams over the same time--frequency
resource~\cite{rusek2013scaling}. Recovering these streams requires
resolving their channel-induced interference. Maximum-likelihood (ML)
detection exploits the finite constellation jointly across all streams, but
its complexity grows exponentially with the number of transmitted
symbols. Linear receivers reduce this complexity at an accuracy penalty
that becomes substantial in highly loaded systems and over ill-conditioned
channels~\cite{yang2015mimo,albreem2019survey}. Efficient detection
therefore requires approximations that preserve useful inter-symbol
dependencies at an affordable computational cost.

Message-passing algorithms address this problem through iterative
approximate inference of the symbol posterior. Approximate message passing (AMP) combines matrix--vector operations with scalar
denoising~\cite{donoho2009amp}; it is asymptotically optimal for large i.i.d.\ Gaussian channels below a
constellation-dependent load threshold~\cite{jeon2015optimality}, but performs poorly on ill-conditioned channels, such as spatially correlated ones~\cite{ma2017oamp}. Expectation propagation (EP) remains robust on such channels at the cost of a matrix inversion per
iteration~\cite{cespedes2014ep,he2024ampgnn}. Both, however, rely on an independent Gaussian approximation of the symbol
posterior that becomes inaccurate under heavy
loading~\cite{kosasih2022gnn}, motivating learned refinement. Unlike deep-unfolded detectors, which are trained for a fixed number of
users~\cite{samuel2019detnet,he2020oampnet}, graph neural networks (GNNs)
match the problem structure: they represent the posterior, a pairwise
Markov random field with channel Gram-matrix couplings~\cite{scotti2020gnn}, with
one node per symbol and a message function shared across all
edges~\cite{kosasih2022gnn}. GEPNet uses such a GNN to refine the EP cavity distributions and
approaches ML performance, but inherits the per-iteration
inversion~\cite{kosasih2022gnn}. GPICNet avoids this inversion by building on Bayesian parallel interference
cancellation, but its accuracy degrades as the number of users
grows~\cite{kosasih2022gnn}. AMP-GNN builds on AMP instead and remains
comparable to GEPNet in fully loaded systems at a much lower complexity~\cite{he2024ampgnn}.

Without the inversion, the cost is dominated by the learned pairwise messages.
With $K$ real symbol components, the AMP-GNN message multilayer perceptron (MLP)
is evaluated on all $K(K-1)$ directed pairs in every graph round; in a
$64\times64$ MIMO system, these evaluations account for 96\% of the
multiplications per received vector. GCEPNet reduces this cost by replacing
the edge-message aggregation with an attention-weighted Chebyshev graph
convolution, but retains EP and its per-iteration
inversion~\cite{lu2024gcepnet}.

Controlling neighborhood size is well established in graph learning; for
example, GraphSAGE samples a fixed number of neighbors to bound its
aggregation workload~\cite{hamilton2017graphsage}. In MIMO detection, EGNN drops a fixed number of weak edges~\cite{li2022egnn},
so its message count still grows quadratically with the system size. Zhou \textit{et al.} prune GNN edges by thresholding EP covariance
correlations~\cite{zhou2023gnnep}, which retains the repeated EP matrix
inversions and leaves the number of neighbors per node uncontrolled.
These approaches motivate a fixed per-node message budget within AMP-GNN,
while keeping its learned nonlinear message for every retained pair.

We first develop a sparse GNN detector (SGNN) in which each symbol component
keeps only the $\kappa$ neighbors with the largest coupling magnitudes. The
message MLP is evaluated only on these retained pairs, reducing the
nonlinear message evaluations per round from $K(K-1)$ to $K\kappa$, which
grows linearly with $K$ for fixed $\kappa$. However, individual coupling
magnitudes do not determine the collective contribution of discarded
interactions to weak channel directions (eigenvectors of the Gram matrix
with small eigenvalues). Aggressive pruning can therefore remove information
that matters for symbol inference even when individual omitted couplings are small.

Pruning also makes the graph local: information from distant nodes must pass
through a few neighbors and is compressed into fixed-size node states, a
limitation known as over-squashing~\cite{alon2021bottleneck}. It is commonly
addressed by rewiring the graph~\cite{giraldo2023tradeoff} or by adding
global communication~\cite{alon2021bottleneck}. Gilmer \textit{et~al.} use a
master node connected to all graph nodes as a global scratch space that lets
information travel long distances~\cite{gilmer2017mpnn}. Southern
\textit{et al.} show that the benefit of virtual nodes depends on the graph
topology and propose a variant that weights nodes according to the graph
structure~\cite{southern2025virtual}. In MIMO detection, RGANet augments
dense local messages with a gated global mean of all node states~\cite{li2025rganet}. In these designs, the global weights are learned or uniform, whereas the channel Gram matrix directly identifies the weak
directions.

We therefore propose a sparse GNN with spectral virtual nodes (SVGNN),
which adds global mixing to the sparse local messages of SGNN.
Each virtual node corresponds to an eigenvector of the Gram matrix and is
connected to all symbol nodes; together, they apply a scaled regularized
inverse of this matrix to the node states, so that every node receives
global information on the weak directions. The virtual nodes add no
learned messages, so the message count remains linear, and their weights
are computed once per channel realization rather than per iteration as in
EP.

Simulation results for different QAM modulation orders over i.i.d.\ and
spatially correlated channels show that SVGNN matches or outperforms
AMP-GNN with substantially fewer multiplications and performs on par with
or better than EP. The virtual nodes improve detection even when added to
the dense AMP-GNN, and in their presence, pruning is essentially lossless.
These results indicate that learned local messages and global
information are complementary: with virtual nodes derived from the system
model, learned messages are needed only between strongly coupled symbols,
which offers a scalable design approach for GNN-based detection.
%%%%%%%%%%%%%%%%%%%%%%%%%%%%%%%%%%%%%%%%%%%%%%%%%%%%%%%%%%%%%%

\textit{Notation:} Boldface lowercase and uppercase letters denote vectors
and matrices, and a tilde marks complex-valued quantities.
$(\cdot)^{\mathsf T}$, $\|\cdot\|$, and $\mathbb E[\cdot]$ denote the
transpose, the Euclidean norm, and the expectation; $\mathbf I_N$ is the $N\times N$
identity matrix; and $\mathcal N$ and $\mathcal{CN}$ denote real and
complex Gaussian distributions.

%%%%%%%%%%%%%%%%%%%%%%%%%%%%%%%%%%%%%%%%%%%%%%%%%%%%%%%%%%%%%%

\begin{figure*}[t]
    \centering
    \includegraphics[width=\textwidth]{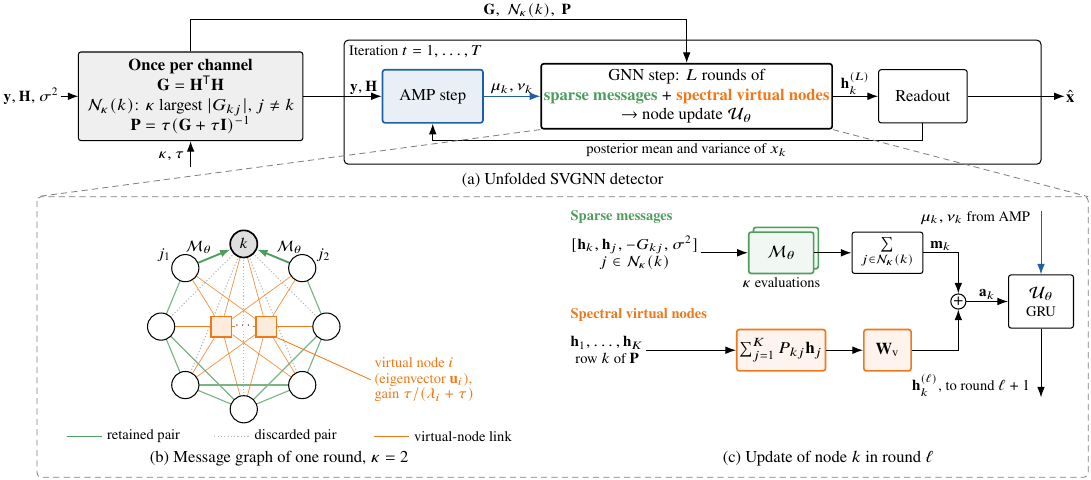}
    \caption{SVGNN detector. (a) The neighbor sets $\mathcal N_\kappa(k)$
    and the matrix $\mathbf P$ are computed once per channel realization from
    the design parameters $\kappa$ and $\tau$. (b) In each GNN round, node $k$
    receives learned messages from its $\kappa$ most strongly coupled
    nodes (green) and is linked to $K$ virtual nodes (orange), one per
    eigenvector of $\mathbf G$. (c) Node $k$ adds the virtual-node term to the sparse aggregate before
    the node update. SGNN is obtained by removing the virtual nodes.}
    \label{fig:architecture}
\end{figure*}

%%%%%%%%%%%%%%%%%%%%%%%%%%%%%%%%%%%%%%%%%%%%%%%%%%%%%%%%%%%%%%
\section{Preliminaries}
\label{sec:preliminaries}

\subsection{System Model}
\label{sec:sysmodel}

We consider uncoded transmission of $N_t$ streams to $N_r$ receive
antennas over a flat-fading channel; the streams may originate from one
multi-antenna transmitter or from $N_t$ single-antenna uplink users. In
the sequel, this configuration is referred to as an $N_t\times N_r$
system. The received signal is

\begin{equation}
    \widetilde{\mathbf y}
    = \widetilde{\mathbf H}\widetilde{\mathbf x}
    + \widetilde{\mathbf n},
    \label{eq:complex_model}
\end{equation}
where $\widetilde{\mathbf H}\in\mathbb C^{N_r\times N_t}$ is the channel
matrix and $\widetilde{\mathbf n}\sim\mathcal{CN}(\mathbf 0,
\widetilde{\sigma}^2\mathbf I_{N_r})$. The entries of
$\widetilde{\mathbf x}$ are drawn independently and uniformly from a
square $M$-QAM constellation with unit average energy, and the SNR is defined as $N_t/(N_r\widetilde\sigma^2)$. We assume that
$\widetilde{\mathbf H}$ and $\widetilde\sigma^2$ are perfectly known at
the receiver.

Two channel models are considered. In the i.i.d. Rayleigh fading channel, the
entries of $\widetilde{\mathbf H}$ are drawn independently from
$\mathcal{CN}(0,1/N_r)$. The correlated channel
follows the Kronecker model
$\widetilde{\mathbf H}=\mathbf R_r^{1/2}\widetilde{\mathbf H}_w
\mathbf R_t^{1/2}\mathbf D$, where $\widetilde{\mathbf H}_w$ is an i.i.d.\
channel, $\mathbf R_r$ and $\mathbf R_t$ are the receive and transmit
exponential correlation matrices with entries $\rho^{|i-j|}$,
$0\le\rho<1$~\cite{loyka2001exp}, and $\mathbf D=\operatorname{diag}(10^{\epsilon_k/20})$
models unequal received powers, with $\epsilon_k$ uniform
in $[-\epsilon_{\max},\epsilon_{\max}]$ dB.

Stacking real and imaginary parts as in~\cite{kosasih2022gnn} gives the real-valued model
\begin{equation}
    \mathbf y = \mathbf H\mathbf x+\mathbf n,\quad
    \mathbf H =
    \begin{bmatrix}
        \Re\{\widetilde{\mathbf H}\}
        & -\Im\{\widetilde{\mathbf H}\}\\
        \Im\{\widetilde{\mathbf H}\}
        & \Re\{\widetilde{\mathbf H}\}
    \end{bmatrix}
    \in\mathbb R^{N\times K},
    \label{eq:real_model}
\end{equation}
where $K=2N_t$, $N=2N_r$, and
$\mathbf n\sim\mathcal N(\mathbf 0,\sigma^2\mathbf I_N)$ with
$\sigma^2=\widetilde{\sigma}^2/2$. Each $x_k$ takes values in the
$\sqrt{M}$-PAM alphabet $\mathcal A$, with $\mathbb E[x_k^2]=1/2$.

With the channel Gram matrix $\mathbf G=\mathbf H^{\mathsf T}\mathbf H$,
the posterior under the uniform prior is
\begin{equation}
    p(\mathbf x\mid\mathbf y,\mathbf H)
    \propto
    \exp\!\left(
        \frac{
            \mathbf y^{\mathsf T}\mathbf H\mathbf x
            -\tfrac12\mathbf x^{\mathsf T}\mathbf G\mathbf x
        }{\sigma^2}
    \right),
    \quad \mathbf x\in\mathcal A^K.
    \label{eq:posterior}
\end{equation}
Its exponent comprises unary terms
$\bigl((\mathbf H^{\mathsf T}\mathbf y)_kx_k-\tfrac12G_{kk}x_k^2\bigr)/\sigma^2$
and pairwise terms $-G_{kj}x_kx_j/\sigma^2$, $k<j$.
Thus, \eqref{eq:posterior} defines a fully connected pairwise
Markov random field with one node per real symbol and
coupling $G_{kj}$ between nodes $k$ and $j$~\cite{scotti2020gnn}.

%%%%%%%%%%%%%%%%%%%%%%%%%%%%%%%%%%%%%%%%%%%%%%%%%%%%%%%%%%%%%%
\subsection{AMP-GNN Detector}
\label{sec:ampgnn}

AMP-GNN~\cite{he2024ampgnn} runs $T$ iterations of an AMP step followed
by a GNN step. At iteration $t$, AMP models each symbol as observed
through a scalar Gaussian channel~\cite{donoho2009amp,jeon2015optimality},
$\mu_k^{(t)}=x_k+w_k^{(t)}$ with
$w_k^{(t)}\sim\mathcal N(0,\nu_k^{(t)})$. The GNN refines these
observations on the graph of~\eqref{eq:posterior} to estimate the symbol
posteriors, where each node $k$ holds a $d$-dimensional hidden state $\mathbf h_k$. In each of the $L$ rounds
of a GNN step, a shared message MLP $\mathcal M_\theta$ computes a
message for every ordered pair $(j,k)$, $j\neq k$, and node $k$ sums its
incoming messages:
\begin{equation}
    \mathbf a_k^{(\ell)}
    =
    \sum_{j\ne k}
    \mathcal M_\theta\!\left(
        [\mathbf h_k^{(\ell-1)},
         \mathbf h_j^{(\ell-1)},
         -G_{kj},\sigma^2]
    \right),
    \label{eq:dense_messages}
\end{equation}
where the iteration index $t$ is omitted. Each node then combines its
aggregate with its AMP observation,
\begin{equation}
    \mathbf h_k^{(\ell)}
    =
    \mathcal U_\theta\!\left(
        \mathbf a_k^{(\ell)},\mu_k,\nu_k
    \right),
    \label{eq:node_update}
\end{equation}
where $\mathcal U_\theta$ is a shared gated recurrent unit (GRU) followed
by a linear layer. After the last round, a shared readout network maps
$\mathbf h_k^{(L)}$ to a probability vector over $\mathcal A$, whose
mean and variance are returned to AMP; after the last iteration, the
means are quantized to give the detected symbols; see~\cite{he2024ampgnn} for the complete
algorithm.

Since~\eqref{eq:dense_messages} is evaluated for all $K(K-1)$ ordered
pairs in every round, the message MLP runs $TLK(K-1)$ times per received
vector. For fixed network widths, this cost grows quadratically with $K$
and dominates the detector complexity (Section~\ref{sec:complexity}).

%%%%%%%%%%%%%%%%%%%%%%%%%%%%%%%%%%%%%%%%%%%%%%%%%%%%%%%%%%%%%%

\section{Proposed SGNN and SVGNN Detectors}
\label{sec:proposed}

Fig.~\ref{fig:architecture} illustrates the proposed detectors. In SGNN,
each node aggregates learned messages only from its $\kappa$ most strongly
coupled neighbors; SVGNN additionally supplies every node with channel-wide
information through spectral virtual nodes.

\subsection{SGNN: Top-$\kappa$ Message Pruning}
\label{sec:pruning}

For each node $k$, let $\mathcal N_\kappa(k)$ be the set of the $\kappa$
indices $j\neq k$ with the largest coupling magnitudes $|G_{kj}|$. Since
$G_{kj}$ is the coefficient of the pairwise term that couples $x_k$ and
$x_j$ in~\eqref{eq:posterior}, this selection retains the strongest direct
interactions of each node. The message MLP is evaluated only on these pairs, and the
dense aggregate~\eqref{eq:dense_messages} is replaced by
\begin{equation}
    \mathbf m_k^{(\ell)}
    =
    \sum_{j\in\mathcal N_\kappa(k)}
    \mathcal M_\theta\!\left(
        [\mathbf h_k^{(\ell-1)},
         \mathbf h_j^{(\ell-1)},
         -G_{kj},\sigma^2]
    \right).
    \label{eq:sparse_messages}
\end{equation}
The sets $\mathcal N_\kappa(k)$ depend only on $\mathbf G$ and are
therefore reused in all $TL$ rounds, so the message MLP is evaluated
$TLK\kappa$ times per received vector, which is linear in $K$. We refer to
this detector as SGNN. A relabeling of the symbols permutes $\mathbf G$,
and hence the sets $\mathcal N_\kappa(k)$, consistently, so SGNN retains
the permutation equivariance of AMP-GNN~\cite{he2024ampgnn}.

\subsection{Limitations of Local Message Passing}
\label{sec:diagnosis}

Consider the spectral decomposition
$\mathbf G=\sum_{i=1}^{K}\lambda_i\mathbf u_i\mathbf u_i^{\mathsf T}$,
where the eigenvectors are orthonormal and
$0\leq\lambda_1\leq\cdots\leq\lambda_K$. Since
$\|\mathbf H\mathbf u_i\|^2=\mathbf u_i^{\mathsf T}\mathbf G\mathbf u_i
=\lambda_i$, perturbations along directions associated with small
eigenvalues produce only small changes in the received signal. These weak
directions are therefore poorly observed by the channel.

To examine how pruning affects these directions, write
\begin{equation}
    \lambda_i
    =\mathbf u_i^{\mathsf T}\mathbf G\mathbf u_i
    =\underbrace{\sum_{k=1}^{K}G_{kk}u_{ik}^2}
        _{\text{diagonal contribution}}
    +\underbrace{2\sum_{k<j}G_{kj}u_{ik}u_{ij}}
        _{\text{coupling contribution}},
    \label{eq:rayleigh}
\end{equation}
where $u_{ik}$ denotes the $k$-th entry of $\mathbf u_i$. Since
$\sum_k u_{ik}^2=1$, the diagonal contribution is a weighted average of the
column energies $G_{kk}$ of $\mathbf H$ and is therefore at least
$\min_k G_{kk}$. An eigenvalue below the smallest column energy can thus
arise only through cancellation by the coupling contribution, which depends
on the signed aggregate of all couplings rather than on the strongest ones
alone.

Let $\mathbf G_\kappa$ retain the diagonal of $\mathbf G$ and each coupling
$G_{kj}$ for which $j\in\mathcal N_\kappa(k)$ or $k\in\mathcal N_\kappa(j)$,
with all other entries set to zero, and define the discarded-coupling
matrix $\mathbf D_\kappa=\mathbf G-\mathbf G_\kappa$. We refer to
$\mathbf u_i^{\mathsf T}\mathbf A\mathbf u_i$ as the directional gain of a
matrix $\mathbf A$ along $\mathbf u_i$. For $\mathbf A=\mathbf G$, it equals
$\lambda_i$, the received energy of a unit perturbation along
$\mathbf u_i$, whereas for the pruned matrix

\begin{figure}[t]
    \centering
    \includegraphics[width=\columnwidth]{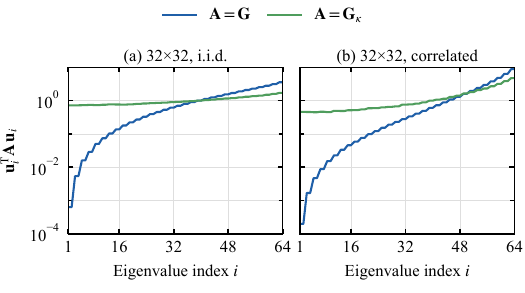}
    \caption{Directional gain $\mathbf u_i^{\mathsf T}\mathbf A\mathbf u_i$
    of the eigenvectors of $\mathbf G$, sorted by eigenvalue, for the full
    ($\mathbf A=\mathbf G$) and pruned ($\mathbf A=\mathbf G_\kappa$,
    $\kappa=2$) graphs in $32\times32$ systems; median over 200 channel
    realizations.}
    \label{fig:spectrum}
\end{figure}

\begin{equation}
    \mathbf u_i^{\mathsf T}\mathbf G_\kappa\mathbf u_i
    =\lambda_i-\mathbf u_i^{\mathsf T}\mathbf D_\kappa\mathbf u_i.
    \label{eq:pruned_direction}
\end{equation}
If the discarded couplings contribute negatively along $\mathbf u_i$,
pruning increases the gain, and for a small $\lambda_i$, even a moderate
increase is a large relative error. 

Fig.~\ref{fig:spectrum} shows the true gain $\lambda_i$ and the pruned
gain for each eigenvector. Along the strongest directions, the two gains
remain close. As $\lambda_i$ decreases, however, the true gain drops by several
orders of magnitude, whereas the pruned gain remains nearly constant over
i.i.d.\ channels and decreases only slightly over correlated channels.
The pruned graph therefore treats precisely the poorly observed
directions as well observed.

These observations call for a global operator that is computed from the
full $\mathbf G$ and emphasizes its weak directions. We use the scaled
regularized inverse
\begin{equation}
    \mathbf P
    =\tau(\mathbf G+\tau\mathbf I)^{-1}
    =\sum_{i=1}^{K}
      \frac{\tau}{\lambda_i+\tau}
      \mathbf u_i\mathbf u_i^{\mathsf T},
    \qquad \tau>0,
    \label{eq:weak_direction_filter}
\end{equation}
whose factors $\tau/(\lambda_i+\tau)$ are close to one for
$\lambda_i\ll\tau$ and small for $\lambda_i\gg\tau$, so the
regularization parameter $\tau$ sets the threshold below which a direction
counts as weak. For $\tau=\sigma^2/E_s$, with $E_s=\mathbb{E}[x_k^2]$, the
linear minimum mean-square error (LMMSE) estimate of $\mathbf x$ is
$\hat{\mathbf x}=(\mathbf G+\tau\mathbf I)^{-1}\mathbf H^{\mathrm T}\mathbf y$,
whose conditional bias is
$\mathbb{E}[\hat{\mathbf x}\,|\,\mathbf x]-\mathbf x=-\mathbf P\mathbf x$.
Thus, $\mathbf P$ emphasizes the directions that linear estimation
suppresses most strongly. In SVGNN, $\tau$ is a design parameter set to a few times $\sigma^2/E_s$,
with a lower bound that keeps $\mathbf P$ from vanishing at high SNR (Section~\ref{sec:setup}).

Even without pruning, the near-cancellation between the diagonal and
coupling contributions in~\eqref{eq:rayleigh} can amplify approximation
errors in the learned messages. If each term $2G_{kj}u_{ik}u_{ij}$ of the
coupling contribution has a relative error of at most~$\epsilon$, the
worst-case relative change in the gain along $\mathbf{u}_i$ is
$\epsilon S_i/\lambda_i$, where
$S_i = 2\sum_{k<j}|G_{kj}u_{ik}u_{ij}| \geq \min_k G_{kk} - \lambda_i$,
so $S_i/\lambda_i$ is large when $\lambda_i \ll \min_k G_{kk}$.
Spectral virtual nodes therefore apply $\mathbf{P}$, providing the
weak-direction weighting directly from $\mathbf{G}$ without learning it.

\subsection{SVGNN: Global Mixing via Spectral Virtual Nodes}
\label{sec:vn}

In every GNN round, SVGNN adds a $\mathbf P$-weighted combination of all
node states to the sparse aggregate,
\begin{equation}
    \mathbf a_k^{(\ell)}
    =
    \mathbf m_k^{(\ell)}
    +\mathbf W_{\mathrm v}\sum_{j=1}^{K}P_{kj}\,\mathbf h_j^{(\ell-1)},
    \label{eq:vn}
\end{equation}
which then enters the node update~\eqref{eq:node_update}. Here, $P_{kj}$
are the entries of $\mathbf P$ in~\eqref{eq:weak_direction_filter}, and
$\mathbf W_{\mathrm v}\in\mathbb R^{d\times d}$ is a weight matrix learned
jointly with the other GNN parameters.
By~\eqref{eq:weak_direction_filter}, the sum in~\eqref{eq:vn} can be written as
\begin{equation}
    \sum_{j=1}^{K} P_{kj}\,\mathbf{h}_j^{(\ell-1)}
    = \sum_{i=1}^{K} \frac{\tau}{\lambda_i+\tau}\, u_{ik}
      \sum_{j=1}^{K} u_{ij}\,\mathbf{h}_j^{(\ell-1)}.
    \label{eq:vn_spectral}
\end{equation}
Each eigenvector thus acts as a virtual node~\cite{gilmer2017mpnn} that
pools the states of all symbol nodes with weights $u_{ij}$ and returns the
pooled state to node $k$ with weight $u_{ik}\tau/(\lambda_i+\tau)$, as
illustrated in Fig.~\ref{fig:architecture}(b). Unlike learned or uniformly
weighted global nodes~\cite{southern2025virtual,li2025rganet}, these
weights are fixed by the channel, and the virtual nodes add only the $d^2$
parameters of $\mathbf W_{\mathrm v}$. Computing $\mathbf P$ requires one $K \times K$ matrix inversion per channel
realization. Since a relabeling of the
symbols permutes $\mathbf P$ in the same way as $\mathbf G$, SVGNN is also
permutation equivariant.

\subsection{Complexity Analysis}
\label{sec:complexity}

We measure complexity by the number of real multiplications required to
detect one received vector $\mathbf y$. The Gram matrix
$\mathbf G=\mathbf H^{\mathsf T}\mathbf H$ is common to all detectors and
is not counted. The neighbor sets and $\mathbf P$, which are computed once
per channel realization, are counted in full for every received vector,
which is the least favorable case. With $C_{\mathcal M}$ and $C_{\mathcal U}$ denoting the multiplications per message evaluation and per node update, the computational complexities are
\begin{equation}
\begin{aligned}
     &\text{AMP-GNN:} && \mathcal O\bigl(TNK+TLK(KC_{\mathcal M}+C_{\mathcal U})\bigr),\\
    &\text{SGNN:}     && \mathcal O\bigl(TNK+TLK(\kappa C_{\mathcal M}+C_{\mathcal U})\bigr),\\
    &\text{SVGNN:}    && \mathcal O\bigl(K^3+TNK+TLK(\kappa C_{\mathcal M}+C_{\mathcal U}+Kd)\bigr),\\
    &\text{EP:}       && \mathcal O\bigl(TK^3\bigr).
\end{aligned}
\label{eq:complexity}
\end{equation}
Here, $TNK$ accounts for the AMP step. Pruning makes the message term
linear in $K$, and the virtual nodes add a quadratic term with
$d\ll C_{\mathcal M}$ and one $K\times K$ inversion per channel
realization, whereas EP inverts a $K\times K$ matrix in every iteration
for every received vector. Table~\ref{tab:complexity} gives the exact counts. Compared with AMP-GNN,
SVGNN reduces the number of multiplications by a factor that grows with
the system size, from 5.3 at $16\times16$ to 18.1 at $64\times64$, where
the virtual nodes account for 7.6\% of the SVGNN cost.

\begin{table}[t]
    \centering
    \caption{Real multiplications per received vector
    ($\times10^6$) for 16-QAM with $T=10$, $L=2$, and $\kappa=2$}
    \label{tab:complexity}
    \begin{tabular}{@{}lccc@{}}
        \toprule
        Detector & $16\times16$ & $32\times32$ & $64\times64$\\
        \midrule
        AMP-GNN  & 78.9 & 299.4 & 1165.4\\
        SGNN     & 14.7 & 29.6  & 59.5\\
        SVGNN    & 15.0 & 30.6  & 64.4\\
        \bottomrule
    \end{tabular}
\end{table}

%%%%%%%%%%%%%%%%%%%%%%%%%%%%%%%%%%%%%%%%%%%%%%%%%%%%%%%%%%%%%%
\begin{figure*}[t]
    \centering
    \includegraphics[width=\textwidth]{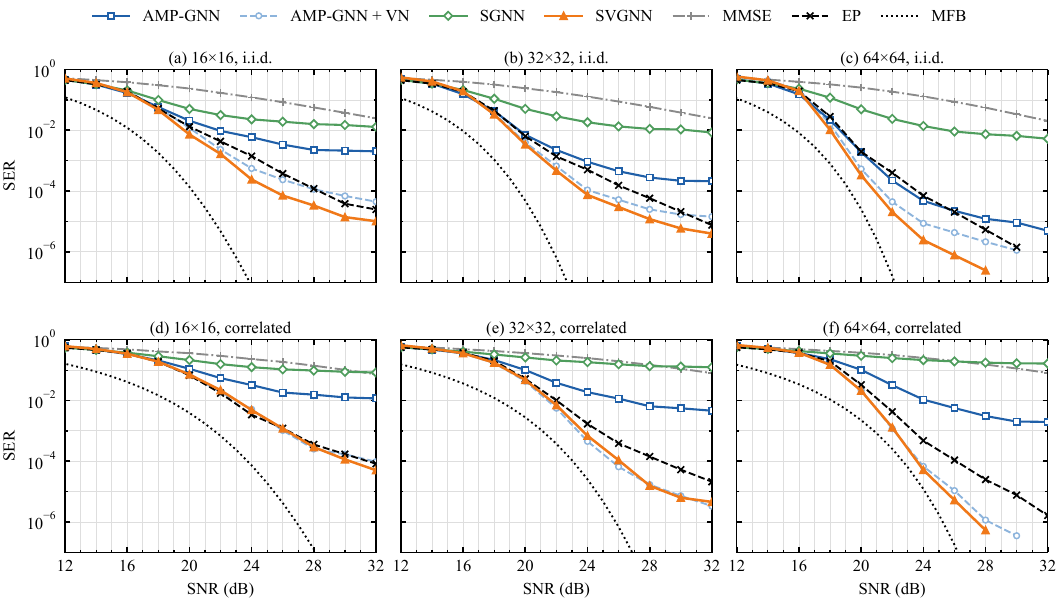}
    \caption{SER of 16-QAM in $16\times16$, $32\times32$, and $64\times64$
    systems over i.i.d.\ (top) and correlated (bottom) channels, with
    $\kappa=2$. AMP-GNN + VN denotes AMP-GNN with spectral virtual nodes,
    and MFB denotes the matched-filter bound.}
    \label{fig:ser16}
\end{figure*}
%%%%%%%%%%%%%%%%%%%%%%%%%%%%%%%%%%%%%%%%%%%%%%%%%%%%%%%%%%%%%%

\section{Simulation Results}

\subsection{Simulation Setup}
\label{sec:setup}

We consider $16\times16$, $32\times32$, and $64\times64$ systems for QPSK
and 16-QAM, and $16\times32$ and $32\times64$ systems for 64-QAM, since
AMP-based detection of higher-order constellations requires more receive
antennas per transmit antenna~\cite{jeon2015optimality}. The correlated
channel of Section~\ref{sec:sysmodel} uses $\rho=0.5$ and
$\epsilon_{\max}=5$~dB.

All learned detectors use $T=10$, $L=2$, $d=8$, a message MLP with layers
of sizes 18, 64, 32, and 8 and ReLU activations, a GRU with 64 hidden
units, and a damping factor of 0.7 in the AMP step. Each detector is trained separately for each configuration over 500 epochs of $10^5$ newly generated samples, with the SNR drawn uniformly in $[12,26]$~dB for QPSK
and $[18,32]$~dB for 16-QAM and 64-QAM. SGNN and SVGNN use $\kappa=2$, and the virtual nodes use
$\tau=4\max(\sigma^2,\sigma_{\mathrm{fl}}^2)/E_s$, where
$\sigma_{\mathrm{fl}}^2$ is the noise variance at an SNR 8~dB above the
lowest training SNR. The loss is the cross-entropy of the symbol
probabilities plus the squared error of the posterior means, minimized by
Adam with an initial learning rate of $2\times10^{-3}$, reduced when the
validation loss plateaus, and batch sizes of 384 and 96 for $N_r\le32$ and
$N_r=64$, respectively. Each curve is averaged over at least four
independently trained models, with at least 200 symbol errors per point.
As references, we use the MMSE detector~\cite{yang2015mimo}, EP~\cite{cespedes2014ep} with $T=10$ iterations, and the matched-filter bound (MFB), that is, the SER obtained when the
interference from all other symbols is perfectly removed. All SNR
differences below are quoted at an SER of $10^{-4}$.

\begin{figure}[t]
    \centering
    \includegraphics[width=\columnwidth]{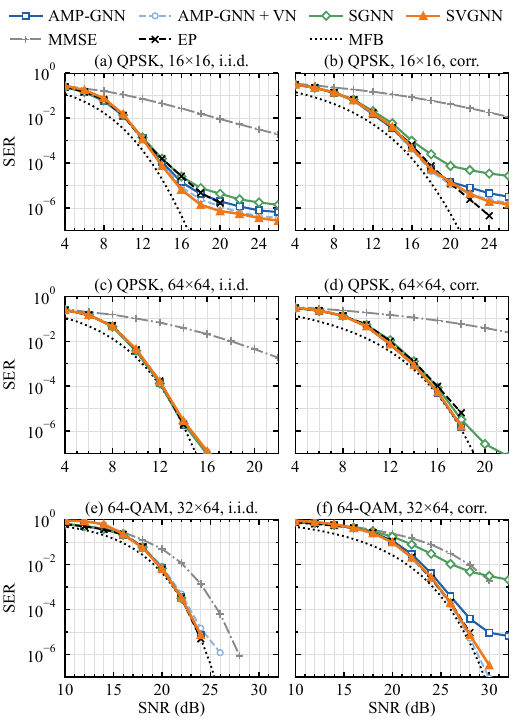}
    \caption{SER of QPSK (a)--(d) and 64-QAM (e), (f) over i.i.d.\ (left)
    and correlated (right) channels, with $\kappa=2$.}
    \label{fig:serqpsk64}
\end{figure}

\subsection{SER Performance}
\label{sec:ser}

Figs.~\ref{fig:ser16} and~\ref{fig:serqpsk64} show that SVGNN matches or
outperforms AMP-GNN in every configuration. For QPSK, the two perform on par,
close to the MFB. For 16-QAM, SVGNN reaches $10^{-4}$ in all configurations,
whereas AMP-GNN exhibits an error floor above this level in all but the
$64\times64$ i.i.d.\ channel, where SVGNN gains 2.2~dB. For 64-QAM, the two coincide on i.i.d.\ channels, whereas on correlated
channels SVGNN gains 0.5 to 0.8~dB and avoids the error floor that AMP-GNN
exhibits.

Compared with EP, SVGNN gains up to 0.7~dB for QPSK and 1.1 to 3.2~dB for
16-QAM, and performs on par for 64-QAM. These gains under full load
indicate that the learned message passing compensates for the inaccuracy
of the Gaussian posterior approximation on which EP
relies~\cite{kosasih2022gnn}. The configurations not shown, $32\times32$
for QPSK and $16\times32$ for 64-QAM, follow the same trends.

\subsection{Effect of Pruning and Spectral Virtual Nodes}
\label{sec:ablation}

To separate the effects of pruning and of the virtual nodes,
Figs.~\ref{fig:ser16} and~\ref{fig:serqpsk64} also show AMP-GNN + VN, that
is, AMP-GNN with spectral virtual nodes but without pruning. For QPSK,
AMP-GNN + VN performs on par with AMP-GNN, whereas for the higher-order
constellations it substantially lowers the error floors of AMP-GNN. Since both detectors operate on the dense graph, this improvement confirms
the analysis of Section~\ref{sec:diagnosis}: the learned messages alone do
not fully reproduce the information that $\mathbf P$ carries about the weak
channel directions.

Without the virtual nodes, pruning is sensitive to the channel and the
constellation. For QPSK, SGNN remains within 0.4~dB of AMP-GNN on i.i.d.\
channels but loses up to 2.0~dB on correlated channels. For 16-QAM, it
exhibits an error floor above $5\times10^{-3}$ in all configurations. For
64-QAM, it coincides with AMP-GNN on i.i.d.\ channels, where the additional
receive antennas keep $\mathbf G$ well conditioned, but saturates above
$10^{-3}$ on correlated channels. Two factors explain this degradation. First, correlation lowers most eigenvalues of $\mathbf G$
(Fig.~\ref{fig:spectrum}), which increases the number of weak directions
that pruning misrepresents.
Second, the tolerance to pruning decreases with the constellation order:
on i.i.d.\ channels, QPSK is largely unaffected, whereas 16-QAM exhibits an
error floor.

Adding the virtual nodes to SGNN removes this degradation: SVGNN matches or
outperforms its dense counterpart AMP-GNN + VN in every configuration, by
up to 3.1~dB. Hence, once the weak directions are supplied globally, the
couplings discarded by pruning are no longer needed. The advantage over
AMP-GNN + VN further suggests that restricting the learned messages to the
strongest couplings, while the virtual nodes carry the global information,
facilitates training.

%%%%%%%%%%%%%%%%%%%%%%%%%%%%%%%%%%%%%%%%%%%%%%%%%%%%%%%%%%%%%%
\section{Conclusion}
\label{sec:conclusion}

To address the quadratic cost of learned message passing in GNN-based MIMO detection, this paper proposed SVGNN, a GNN detector that combines sparse local messages with spectral virtual nodes. The sparse messages restrict the learned interactions of each symbol to its most strongly coupled neighbors, so that their number grows linearly with the system size. The virtual nodes, derived from a regularized inverse of the Gram matrix, supply every node with the weak channel directions, which coupling magnitudes do not reveal and which learned messages capture only partially, even on the dense graph. Since the virtual nodes add no learned messages, SVGNN requires only a fraction of the multiplications of AMP-GNN. Simulation results demonstrated that, despite this reduction, SVGNN matches or outperforms AMP-GNN and EP across constellations, system sizes, and channel conditions. These findings indicate that a GNN detector need not learn every interaction: when an operator derived from the system model provides the global structure of the inference problem, learning can be confined to local interactions.

\bibliographystyle{IEEEtran}
\bibliography{refs}

\end{document}